\documentclass[aps,prl,reprint,superscriptaddress]{revtex4-2}
\usepackage{graphicx}
\usepackage[separate-uncertainty=false]{siunitx}
\usepackage{lineno}
\usepackage{framed}
\usepackage{ragged2e}
\usepackage{multirow}
\usepackage{amsmath}
\graphicspath{figures/}
\usepackage{lineno}

\begin{document}

\title{Generation and detection of squeezed light on a single silicon photonic chip}
\author{Oliver~M.~Green}
\affiliation{Quantum Engineering Technology Labs, H. H. Wills Physics Laboratory and School of Electrical, Electronic, and Mechanical Engineering, University of Bristol, BS8 1FD, UK}
\affiliation{Quantum Engineering Centre for Doctoral Training, H. H. Wills Physics Laboratory and School of Electrical, Electronic, and Mechanical Engineering, University of Bristol, BS8 1FD, UK}

\author{Bethany~Puzio}
\affiliation{Quantum Engineering Technology Labs, H. H. Wills Physics Laboratory and School of Electrical, Electronic, and Mechanical Engineering, University of Bristol, BS8 1FD, UK}
\affiliation{Quantum Engineering Centre for Doctoral Training, H. H. Wills Physics Laboratory and School of Electrical, Electronic, and Mechanical Engineering, University of Bristol, BS8 1FD, UK}

\author{Rowan~A.~Hoggarth}
\affiliation{Quantum Engineering Technology Labs, H. H. Wills Physics Laboratory and School of Electrical, Electronic, and Mechanical Engineering, University of Bristol, BS8 1FD, UK}

\author{Rachel~N.~Clark}
\affiliation{Quantum Engineering Technology Labs, H. H. Wills Physics Laboratory and School of Electrical, Electronic, and Mechanical Engineering, University of Bristol, BS8 1FD, UK}

\author{Edward~C.~R.~Deacon}
\affiliation{Quantum Engineering Technology Labs, H. H. Wills Physics Laboratory and School of Electrical, Electronic, and Mechanical Engineering, University of Bristol, BS8 1FD, UK}

\author{Alex~S.~Clark}
\affiliation{Quantum Engineering Technology Labs, H. H. Wills Physics Laboratory and School of Electrical, Electronic, and Mechanical Engineering, University of Bristol, BS8 1FD, UK}

\author{Jonathan~C.~F.~Matthews}
\affiliation{Quantum Engineering Technology Labs, H. H. Wills Physics Laboratory and School of Electrical, Electronic, and Mechanical Engineering, University of Bristol, BS8 1FD, UK}

\author{Giacomo~Ferranti}
\email[]{gcm.ferranti@gmail.com}
\affiliation{Quantum Engineering Technology Labs, H. H. Wills Physics Laboratory and School of Electrical, Electronic, and Mechanical Engineering, University of Bristol, BS8 1FD, UK}
\date{\today}

\begin{abstract}
The ability to generate and detect quantum states of light on a single integrated photonic device is essential to scale quantum photonics into useful quantum technologies. Integrating the required capabilities into complementary-metal-oxide-semiconductor compatible monolithic chips can reduce cost and unlock new functionality through miniaturisation. In this work we demonstrate a single silicon-on-insulator photonic integrated circuit for the monolithic generation and detection of quantum light on a commercially available platform that operates entirely at room temperature. Specifically, we leverage spontaneous four-wave mixing in silicon waveguides to produce squeezed light which is subsequently detected by photodiodes operating in a pulsed homodyne detector configuration on the same chip as the source. We directly measure $\SI{0.25(1)}{\decibel}$ of squeezing, including contributions from waveguide propagation loss and detection inefficiency, and include a detailed analysis of the impact of nonlinear loss on the squeezing levels achievable using this platform. 
\end{abstract}


\maketitle

\section{Introduction}
\label{sec:intro}
Integrated photonics has been revolutionising the development of miniaturised and scalable optical quantum technologies for several years~\cite{wang_integrated_2020,moody_2022_2022}. Photonic integrated circuits (PICs) have been employed extensively in quantum simulation \cite{peruzzo_variational_2014,somhorst_quantum_2023}, communication \cite{luo_recent_2023}, sensing and metrology \cite{labonte2024integrated} due to their compactness, stability and scalability. In the race towards large scale quantum photonic applications, monolithic integration of all required photonic components onto a single photonic platform compatible with complementary metal-oxide-semiconductor (CMOS) processes is often identified as a necessary step. Such developments have been pursued in discrete variable quantum photonics where waveguide coupled superconducting nanowire single photon detectors (SNSPDs) have been co-integrated with both quantum dot~\cite{reithmaier_-chip_2013} and heralded parametric single photon sources~\cite{alexander_manufacturable_2025}. In contrast, no such demonstration has been performed within the continuous variable (CV) paradigm where balanced homodyne detectors~\cite{kumar_versatile_2012} replace SNSPDs for state measurement. Ubiquitous across CV quantum technology is the use of squeezed light which is readily and deterministically generated at room temperature through nonlinear optical processes. Squeezed light has been leveraged as a core component in CV quantum computing architectures~\cite{menicucci_universal_2006,bourassa_blueprint_2021,aghaee_rad_scaling_2025} and sensing applications~\cite{qin_unconditional_2023,giovannetti_quantum_2006}, including in enhanced gravitational wave detection~\cite{the_ligo_scientific_collaboration_gravitational_2011,vbirgo_2019}, absorption~\cite{atkinson_quantum_2021}, biological imaging~\cite{taylor_subdiffraction-limited_2014} and microscopy~\cite{casacio_2021}. As such, a large effort has been made to develop components for CV quantum technologies on PICs~\cite{clark_integrated_2026}.

Shot-noise-limited homodyne detectors on silicon-on-insulator (SOI) PICs have been demonstrated using waveguide coupled photodiodes~\cite{raffaelli_homodyne_2018}. Since these photodiodes offer simultaneously high bandwidth and responsivity, the last few years have seen silicon-integrated germanium photodiode based homodyne detectors pushing the boundary of achievable homodyne detection bandwidth~\cite{tasker_silicon_2021, bruynsteen_integrated_2021, tasker_bi-cmos_2024}. 

In contrast to the detectors, typical integrated sources of squeezed light are based on silicon nitride (SiN)~\cite{dutt_-chip_2015,vaidya_broadband_2020, jahanbozorgi_generation_2023,zhao_near-degenerate_2020, zhang_squeezed_2021, shen_strong_2025,larsen_integrated_2025,jia2026monolithic,Danilin_2026} or lithium niobate ~\cite{chen_ultra-broadband_2022,peace_picosecond_2022,stokowski_integrated_2023,park_single-mode_2024,arge_demonstration_2024} waveguides, both platforms in which quantum compatible homodyne detectors have yet to be demonstrated. These experiments have therefore relied on characterisation of the generated squeezing via external homodyne detectors. This not only impacts the total device footprint, and thus scalability, but incurs detrimental loss on the squeezing. For example, the squeezing detected by Zhang \textit{et al.}~\cite{zhang_squeezed_2021} was limited to $\SI{1.65}{\decibel}$ despite estimated generated levels of $\SI{8}{\decibel}$ with the major loss contribution being that of chip-to-fibre coupling. This, alongside the ability to facilitate large arrays of detectors in a scalable manner~\cite{gurseson_y2025}, motivates the development of a monolithic CV PIC. However, due to this disconnect between the platforms that squeezed light can be generated on and detected in, realising monolithic integration of CV sources and detectors on a single PIC has proved challenging.

Whilst heterogeneous integration of homodyne detectors has been demonstrated though flip chip bonding~\cite{bai_188_2021}, the performance of such devices currently lag behind the state of the art SOI detectors, restricting their efficacy for quantum applications. Likewise, current demonstrations of heterogeneously integrated photodiodes on SiN and lithium niobate exhibit C-band responsivities of \SI{0.8}{\ampere\per\watt} \cite{yu2020heterogeneous,gao2025heterogeneous} and \SI{0.4}{\ampere\per\watt} \cite{wei2023ultra,wei_ultrahigh-speed_2024} respectively, below the \SI{1.1}{\ampere\per\watt} of the SOI detectors used in this work. 
\begin{figure*}
    \centering
    \includegraphics[width = \linewidth]{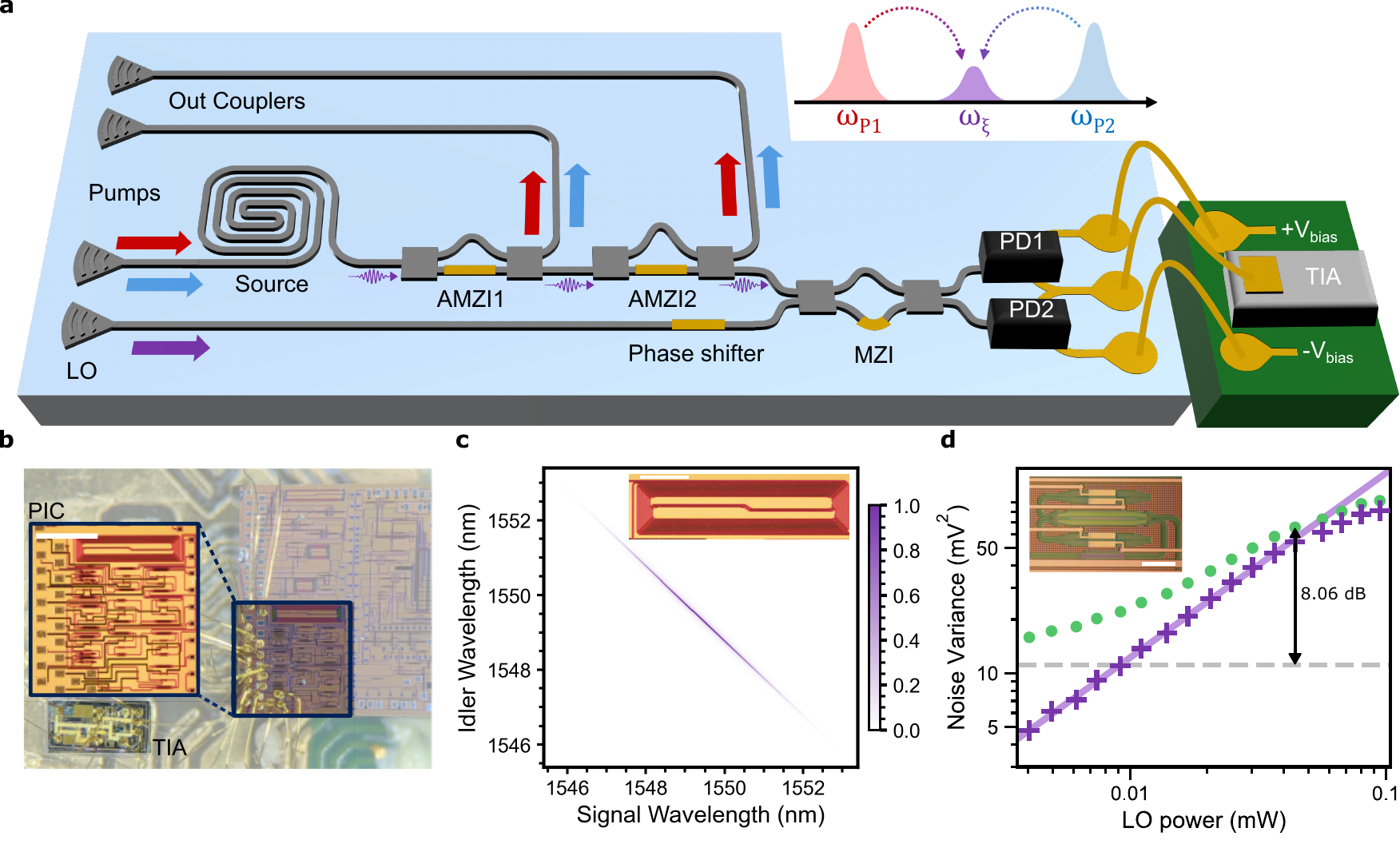}
\caption{\textbf{The photonic chip} (a) Layout of the SOI photonic integrated circuit (PIC). A bi-chromatic pump and local oscillator (LO) couple into the PIC via separate grating couplers. The pump generates degenerate squeezing, as shown in the insert, by spontaneous four-wave mixing in a spiral waveguide. Two cascaded asymmetric Mach Zehnder interferometers (AMZIs) separate squeezed light from pump. The former is then routed to the on-chip homodyne detector, composed of waveguide-coupled germanium photodiodes at each output of a Mach Zehnder interferometer (MZI). The LO field couples directly from its input grating coupler into the homodyne MZI, where it interferes with the squeezed light. (b) Optical microscope image of the PIC wirebonded to the TIA, with high resolution images of the PIC shown in the inset, scale bar \SI{500}{\micro\meter}. (c) The simulated joint spectral intensity for the spiral source with the spectral filtering of the AMZI applied to the broadband phasematching. A close in microscope image of the spiral waveguide source is shown in the inset, scale bar \SI{200}{\micro\metre}. (d) Shot noise clearance of the pulsed homodyne detector. Green circles are the raw noise variances, purple crosses show the noise variance with the noise floor (grey dashed line) subtracted. The purple linear fit gives a gradient of 1.04 and a maximum clearance of \SI{8.06}{\decibel}. A close in microscope image of the homodyne detector is shown in the inset, scale bar \SI{50}{\micro\metre}.}
    \label{fig:1}
\end{figure*}
The SOI platform is currently one of the few commercially available platforms that simultaneously offers nonlinear sources of quantum states and high-performance photodetectors, but squeezed light generated from silicon waveguides had not been directly measured until now, with generated squeezing only being inferred from boson sampling experiments~\cite{paesani_generation_2019}.

Here, we report the generation and homodyne detection of degenerate squeezed vacuum states of light on a single monolithic SOI PIC. The state generation, manipulation, and pulsed homodyne detection were all performed within the same PIC, via integrated photonic structures, without need for off-chip coupling. The PIC has a footprint of $\SI{1.44}{\mm}^{2}$ and operates entirely at room temperature. Moreover, our work provides direct evidence of squeezed light generation from SOI waveguides. 

\section{Results}
\begin{figure*}[!htb]
    \centering
    \includegraphics[width = \linewidth]{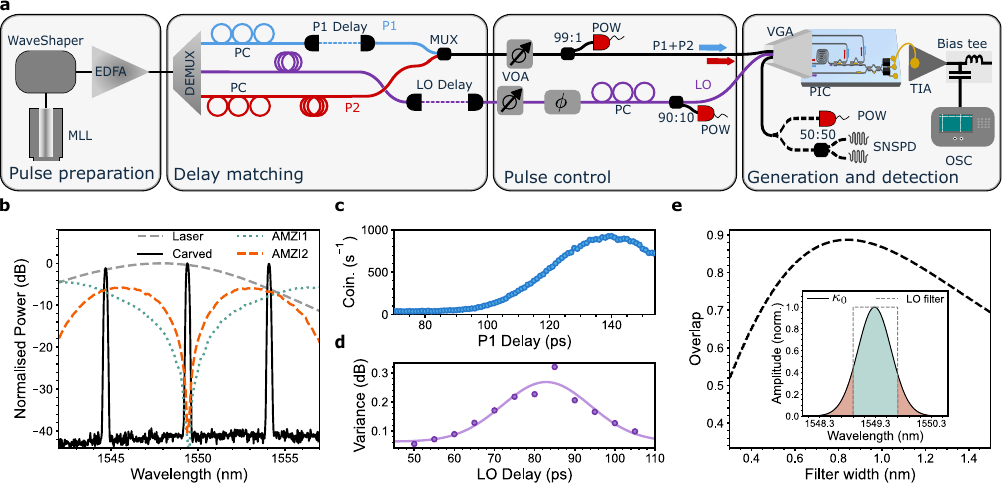}
    \caption{\textbf{Dual-pump preparation scheme} (a) Schematic of the experiment, divided into four stages. Pulse preparation: Three spectral lines carved by the WaveShaper from a spectrally-broad pulsed pump enter an erbium-doped fibre amplifier (EDFA). Delay matching: The three wavelengths are demultiplexed (DEMUX) and tunable delay lines correct the temporal mismatch between them. Pumps P1 and P2 are recombined in a wavelength multiplexer (MUX) to act as a bi-chromatic pump field. A variable optical attenuator (VOA) sets the pump intensity which is monitored by a 99:1 splitter and a power meter (POW). A VOA and phase modulator ($\phi$) are used to control the power and phase of the local oscillator (LO). Squeezing and detection: P1, P2 and LO are coupled onto the chip via a V-groove array (VGA), to perform both the generation and detection of quadrature-squeezed vacuum. The residual outputs of the two on-chip AMZIs are coupled off chip and either measured on power meters or routed to superconducting nanowire single photon detectors (SNSPDs) to perform pump delay matching.  (b) Spectral profiles of the broad pulsed laser, three spectral lines each with bandwidth $\SI{0.2}{\nm}$ carved by the WaveShaper, and the response of the two on-chip AMZIs. (c) Rate of photon coincidences generated by the spiral source as a function of relative delay between P1 and P2. Optimal temporal matching corresponds to maximal photon pair generation. (d) Measured anti-squeezing as a function of the local oscillator delay. Maximum anti-squeezing corresponds to zero relative delay between the generated squeezed light and LO indicating well overlapped temporal modes. (e) Spectral overlap between the measured squeezing (first Schmidt mode, $\kappa_0$) and the LO as a function of the LO top-hat filter width. Inset: the peak-normalised mode profiles of the targeted squeezing spatio-temporal mode, $\kappa_0$, and optimised LO \SI{0.8}{\nano\metre}-wide top-hat filter. Areas in green (red) indicate the spectral regions of squeezing accessible (inaccessible) to the LO.
    }
    \label{fig:2}
\end{figure*}
\subsection{The photonic integrated circuit}
A schematic of the PIC used for simultaneous generation and detection of squeezed light is shown in Fig.~\ref{fig:1}a. The PIC was designed in-house with fabrication performed by IMEC as part of an MPW using their ISiPP50G process with a microscope image shown in Fig.~\ref{fig:1}b. A $\SI{1.1}{\cm}$ spiral waveguide acts as a source \cite{silverstone2014chip}, generating squeezed vacuum via degenerate spontaneous four-wave mixing (SFWM) from a bichromatic pulsed pump coupled onto the chip via a grating coupler. 
As a consequence of the broad satisfaction of the phasematching conditions for the SFWM process in the spiral, the source generates multimode squeezed light which can be expressed as a series of orthogonal modes. To estimate the modal content of the output, we performed a Schmidt decomposition on the joint spectral intensity (JSI) of the output modes of the SFWM process. The calculated JSI of the squeezed vacuum state produced in the spiral waveguide and subsequently filtered by the two AMZIs is shown in Fig.~\ref{fig:1}c from which we determined the Schmidt number to be $K = 34.7$ corresponding to a spectral purity of $2.9\%$. Whilst squeezing is generated in multiple modes, we engineered the LO spectrum to select a single Schmidt mode from the ensemble.
After the spiral waveguide, the squeezed state and the bichromatic pump are guided to two cascaded tunable asymmetric Mach-Zehnder interferometers (AMZIs) that filter away the pump wavelengths. 
Finally, the squeezed light is guided to an on-chip pulsed homodyne detector. The LO couples in via a separate grating coupler, which passes directly to the homodyne detector.
The detector is composed of a symmetric Mach-Zehnder interferometer (MZI), used as a tunable beam splitter, and two waveguide-coupled germanium photodiodes. Both the MZIs and AMZIs are comprised of 50:50 multi-mode interferometer (MMI) splitters and are tuned using thermo-optic phase shifters. The nominal responsivity of the photodiodes at \SI{1550}{\nm} is \SI{1.1(1)}{\ampere\per\watt}, corresponding to a quantum efficiency of $88(8)\%$. The photodiodes are connected in a balanced configuration via the PIC metal routing and the differential photocurrent is output via wirebonds to an unpackaged transimpedance amplifier (TIA), as shown in Fig.~\ref{fig:1}b. This avoids the parasitic capacitance of PCB tracks, allowing high-speed amplification of the homodyne signal \cite{tasker_silicon_2021}. The PIC and TIA are mounted on a custom PCB containing the required electronics for the photodiode biasing and TIA power supply (see Supplementary Information Section III for further details). The mounted chip and optical coupling system are placed inside a Faraday cage to isolate the system from environmental noise. The detector exhibits a \SI{3}{\decibel} bandwidth of \SI{1.56}{\GHz} and a maximum shot noise clearance of \SI{8.06}{\decibel} above the electronic noise floor, shown in Fig.~\ref{fig:1}d, corresponding to a detection efficiency of $83\%$. Due to the high peak power of the pulsed LO field, time-domain pulsed homodyne detection requires careful balancing to prevent saturation due to residual common-mode signals. Therefore, to avoid any risk of detector saturation, we operate at a lower LO power with $\SI{6.67}{\decibel}$ of shot noise clearance. See Supplementary Information Section IV for a full characterisation of the homodyne detector.



\subsection{Generation and control of pump and LO fields}
Since homodyne detection is an interferometric technique, we require the input state and the LO to interfere coherently, meaning that all frequency components of input state and LO must be phase-coherent. This was achieved by spectrally carving the two pump wavelengths and the LO from a single spectrally broad \SI{300}{\fs} mode locked laser (MLL) with a \SI{100}{\MHz} repetition rate. The pulse preparation setup is shown in Fig.~\ref{fig:2}a. The output of the MLL (OneFive Origami 15) was directed into a programmable multi-band optical filter (Finisar WaveShaper 1000A) and carved into three separate spectral lines. Two of these lines, with central wavelengths \SI{1544.7}{\nm} (P1) and \SI{1554.1}{\nm} (P2), were used to pump the SFWM process, whilst the line at \SI{1549.4}{\nm} was used as the LO. These respectively align with standard dense wavelength division multiplexer (DWDM) channels 41, 29 and 35, which enables demultiplexing into separate fibres at a later stage. The WaveShaper applied top-hat filters to define the pumps and LO. The filter for P1 and P2 was set to a spectral bandwidth of \SI{0.2}{\nano\meter} to ensure good extinction by the AMZIs. The \SI{4.7}{\nm} detuning between the pumps and the LO was selected to maximise the outcoupling of the two pumps from the chip by the on chip AMZIs. The overlap of the carved pulses and AMZI spectra are shown in Fig.~\ref{fig:2}b (see Supplementary Information Section II for more details). We determined the optimal LO width by maximising its overlap with the first Schmidt mode ($\kappa_0$), shown in Fig.~\ref{fig:2}e. The inset shows the target spectral profiles for the squeezed vacuum field and the LO, highlighting in green (red) the spectral portions of the former that are accessible (inaccessible) by the homodyne measurement. An overlap of 0.88 between the first Schmidt mode and LO for the optimal filter width of \SI{0.8}{\nano\meter} was calculated, with none of the other first ten Schmidt modes exceeding an overlap of 0.04.

The output of the WaveShaper was amplified by an erbium-doped fibre amplifier (EDFA) (Pritel). Fibre WDMs were used to separate the spectral lines and provide further filtering to suppress the unused parts of the laser spectrum and any amplified spontaneous emission from the EDFA. P1 and P2 were subsequently recombined into a single fibre via a $2\times1$ fibre combiner. The powers of the three fields were controlled by variable optical attenuators (VOAs) and monitored on power meters connected to fibre taps before coupling onto the chip via a V-Groove Array (VGA) aligned to the grating couplers.  Each spectral component had an individual polarisation controller to ensure it is efficiently coupled.

\begin{figure*}[t!]
    \centering
    \includegraphics[width = \linewidth]{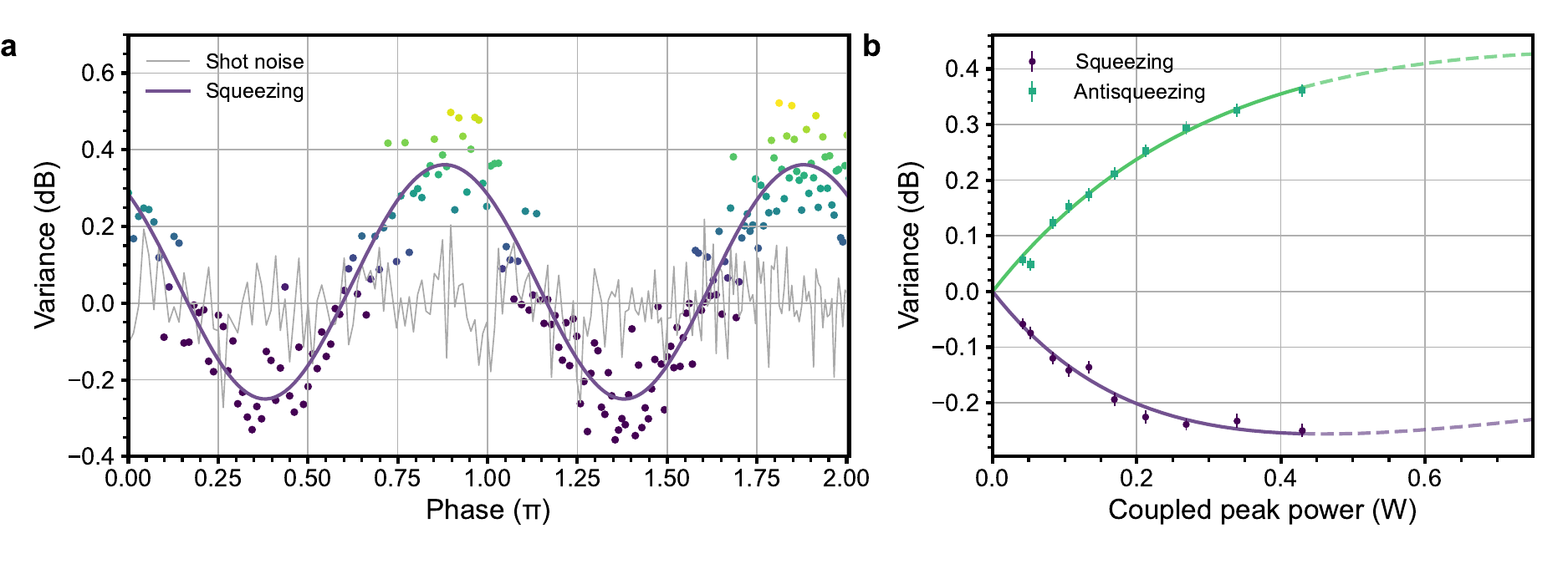}
    \caption{\textbf{Squeezed light measurement} (a) The evolution of the variance of the signal recorded on the integrated homodyne detector as the LO phase is swept, exhibiting clear sub-shot noise fluctuations reaching  $\SI{0.25(1)}{\decibel}$ of squeezing. (b) The dependence of the squeezing (purple circles) and antisqueezing (green squares) generated and detected on chip as peak pump power is varied, reaching a maximum detected squeezing level of $\SI{0.25(1)}{\decibel}$. The data is fitted with the theoretical scaling from Eq.~\ref{power dep}. An extension of the fit to higher optical powers (dashed lines) shows generated squeezing saturates due to the effects of nonlinear loss.}
    \label{fig:3}
\end{figure*}

Propagation through different fibres led to mismatches in the temporal delays of the three optical fields. Compensating these delays was critical to ensure temporal overlap between the pumps in the source and between LO and squeezed light in the homodyne detector. An initial, coarse (to within \SI{1}{\cm}), path length matching operation was performed by measuring each path with an optical back-reflectometer (LUNA OBR) and adding fibre length on the shortest paths. Finer compensation of the relative delays across the three fields was achieved by means of two programmable delay lines placed on the P1 and LO paths, as shown in Fig.~\ref{fig:2}a. Temporal matching between the two pumps was achieved by setting the P1 delay to maximise the rate of photon pairs produced by the SFWM process within the spiral waveguide. The detected coincidence rate with varying delay is shown in Fig.~\ref{fig:2}c. The delay between the LO and squeezed light was compensated by maximising the anti-squeezing signal detected by the homodyne detector, which results in a variance increase compared to shot noise as shown in Fig.~\ref{fig:2}d. More details on these characterisation procedures are included in the Methods section.

\subsection{Squeezing estimation}
Once relative temporal delays had been optimised, we swept the LO phase over a $2\pi$ ramp and recorded the corresponding quadrature measurements (see Supplementary Information Sections V and VI for further details). The variances over intervals of 5000 quadrature measurements were calculated. As these intervals correspond to a phase variation of $0.01$ units of $\pi$, changes in quadrature distribution within them were considered negligible. As shown in Fig.~\ref{fig:3}a, these quadrature variances exhibited the sinusoidal variation with period $\pi$ that is expected from a squeezed vacuum state, with minima below the shot noise level. We extracted the squeezing and anti-squeezing levels by fitting with 
\begin{align}
\label{eqn:fit}
   \Delta X^2\left(\phi\right) = \Delta X^2_{\mathrm{min}} \sin^2\left(\phi\right) +  \Delta X^2_{\mathrm{max}} \cos^2\left(\phi\right), 
\end{align}
where $\Delta X^2\left(\phi\right)$ is the measured variance at phase $\phi$ while $\Delta X^2_{\mathrm{min}}$ and $\Delta X^2_{\mathrm{max}}$ are the measured minimum and maximum variances, respectively. We directly observed a maximum squeezing level of $\SI{0.25(1)}{\decibel}$. The value of squeezing entering the detector is inferred as $\SI{0.42(2)}{\decibel}$ by removing the effect of homodyne detector inefficiency on the measured value (see Supplementary section VII for more information). 
We investigated the dependence of the squeezing level on pump power by altering the pump attenuation and extracting the minimum and maximum variances for each power with Eq.~\ref{eqn:fit}. The results are shown in  Fig.~\ref{fig:3}b. The expected minimum and maximum variances depend on the peak pump power as 
\begin{equation}
\label{power dep}
     \Delta X^2_{\substack{\mathrm{max}\\\mathrm{min}}}\left(P_1,P_2\right)=  1 - \eta + \eta e^{\pm g\left(P_1,P_2\right)},
\end{equation}
where $P_{1,2}$ is the peak power of each pump, $g$ is a power dependent gain parameter and $\eta$ is the overall efficiency that includes contributions from linear and nonlinear loss in the waveguides, non-unity photodetector efficiency and noise contamination in the readout electronics (see Supplementary Section VII for further details). Nonlinear loss mechanisms in silicon waveguides are well documented~\cite{husko2013multi} and are comprised of two-photon absorption (TPA) which we take to be two pump photons being absorbed, cross two-photon absorption (XTPA) which we take to be a pump photon and a generated photon from SFWM being absorbed, and the associated free-carrier absorption (FCA) from TPA and XTPA. These give rise to a power dependence of the efficiency. The choice of a pulsed pump helps to minimise the effect of FCA since the time between pulses is much longer than the $\SI{\sim2}{\nano\second}$ carrier lifetime in silicon waveguides \cite{Pernice_Carrier_and_thermal_2011} and so we neglected this contribution to the efficiency. 
We can therefore write $\eta{\left(p\right)} = \eta_{\mathrm{XTPA}}\left(p\right)\eta_{\mathrm{L}}\eta_{\mathrm{HD}}$ where $\eta_{\mathrm{L}}$ is the transmission accounting for linear propagation loss in the waveguides, $\eta_{\mathrm{HD}}$ is the efficiency associated with the homodyne detector and $p = P_1 + P_2$ is the total peak power in the waveguide. Nonlinear loss on the pumps must also be accounted for in the definition of $g$ where it can be shown that $g$ takes the form \cite{husko2013multi}
\begin{equation}
\label{gamma}
    g = \gamma \frac{A}{\beta_{\mathrm{TPA}}}\ln\left( 1 + \beta_{\mathrm{TPA}}\sqrt{P_1P_2}L/A\right)
\end{equation}
for source length $L$ with waveguide effective area $A$, TPA coefficient $\beta_{\mathrm{TPA}}$ and waveguide nonlinearity $\gamma = {n_2 k_0}/{A}$ where $n_2$ is the nonlinear refractive index and $k_0$ is the average wavevector of the pumps. We experimentally estimated the nonlinear loss coefficient $\alpha_{\mathrm{TPA}} = \beta_{\mathrm{TPA}}/A$, for our source by directing the squeezing to a SNSPD and performing pump power dependent photon counting measurements (see Supplementary Information Section VII for details). We insert Eq.~\ref{gamma} into Eq.~\ref{power dep} and use the experimentally determined $\alpha_{\mathrm{TPA}}$ in $g$ and to fix $\eta_{\mathrm{XTPA}}\left(p\right)$ and then fit the data to determine $\gamma =$~\SI{112.5(56)}{\per\watt\per\meter} and $\eta_{L} = 0.52(6)$. From the value of $\gamma$ we found an estimated nonlinear refractive index of $n_2 = $ \SI{6.38(32)e-18}{\metre\squared\per\watt} which is in good agreement with literature reported values for silicon waveguides \cite{sinclair_temperature_2019}. 

\section{Discussion}
\label{Discussion}
This work demonstrates the generation and characterisation of squeezed light within a single monolithic SOI photonic chip. State generation, manipulation and detection were all performed via integrated photonic structures on the same integrated photonic device on the $\SI{1}{\mm}^{2}$ scale. We directly observed $\SI{0.25(1)}{\decibel}$ of squeezing at the on-chip homodyne detector, which corresponds to an inferred squeezing of $\SI{0.42(2)}{\decibel}$ entering the detector. If we further remove the remaining linear loss contributions experienced by the squeezing through the chip, determined by the fit of Fig.~\ref{fig:3}b, we find a total generated squeezing level of $\SI{0.83(3)}{\decibel}$. 
The device was operated entirely at room temperature, and was based entirely on photonic components available commercially from photonic foundries, which makes its replication more accessible to a wider range of users, removing the need for specialised cleanroom equipment. This demonstrates that SOI remains the main mass manufacturable platform providing both high efficiency photodiodes and nonlinear optical quantum state sources. This work has demonstrated a PIC platform that bridges the gap between platforms for generating squeezed light and performing CV measurements and a route to monolithic CV quantum technologies.

While the focus of this work is on the monolithic integration aspect of the technology, rather than an optimisation of the amount of squeezing produced, we identify the generation of higher squeezing levels as a necessary future step towards practical applications. The most significant hurdle to overcome is the nonlinear loss intrinsic to SOI waveguides when operating in the C-band. This is highlighted by the extension of the fit in Fig.~\ref{fig:3}b to higher powers showing the saturation and decay of the squeezing levels obtained as the power increases. From this we infer that, due to nonlinear loss, measured squeezing levels, on this device would not exceed $\SI{0.26}{\decibel}$. Methods to overcome or mitigate the impact of nonlinear loss on silicon waveguides have been studied at length, including migrating to longer wavelengths towards the two-photon bandgap energy \cite{rosenfeld_mid-infrared_2020} or cooling down the waveguides but therefore compromising on room temperature operation~\cite{sinclair_temperature_2019}. The development of hybrid quantum PICs or the pursuit of heterogeneous integration \cite{bai_188_2021,yu2020heterogeneous,gao2025heterogeneous,wei2023ultra,wei_ultrahigh-speed_2024} techniques across multiple platforms offer further routes to combining sources and detectors, but are yet to reach the commercial maturity of SOI.

A near term application of particular interest is the ability to realise innately scalable quantum sensors. 
Such devices perform quantum-enhanced measurement schemes requiring modest squeezing levels. The small form factor and phase stability offered by a monolithic PIC coupled with room temperature operation offers a route to portable devices. The ability to both generate and detect on the same chip removes the requirement to outcouple the weak quantum signal, demonstrating the robustness and advantages of the fully monolithic approach for enabling deployable and scalable quantum technologies.





\section{Methods}

\subsection{Delay matching of pump and LO fields}
Compensation of the temporal matching between the spectral components of the bichromatic pump within the spiral was performed by maximising the efficiency of the degenerate SFWM process as a function of the delay. We tuned the first AMZI on the output of the spiral waveguide to route the degenerate SFWM field off-chip and into a 50:50 fibre beam splitter leading to two SNSPD channels. A programmable delay placed on the P1 path was swept to find a maximum in the detected photon coincidence rate, which corresponds the optimal temporal overlap of the pumps in the source. The result of this measurement is shown in Fig.~\ref{fig:2}c. We chose to optimise on coincidences rather than raw counts to remove the impact of spurious events from leaked pump light or temporally uncorrelated events such as amplified spontaneous emission.

The relative delay between LO and squeezed light was also fine-tuned by means of a programmable delay, which was placed on the LO path. As a reference signal to optimise the overlap between these two fields, we used direct measurements of anti-squeezing. The output of the homodyne detector was connected to an oscilloscope (Keysight DSOV134A) via a bias tee in order to measure the AC signal.  The LO phase was swept using an external fibre-coupled lithium niobate phase modulator (Thorlabs) to implement a phase rotation of approximately $2\pi$ in 10 ms. A single quadrature was extracted from each of the $10^6$ pulses detected in this time and normalised to the shot noise level (see Supplementary Information Section VI for further details). These quadratures were grouped into bins of 5000 and calculated the variance of each bin. This procedure was repeated for different settings of the programmable delay, recording the maximum variance obtained within the phase ramp for each delay. The results are plotted in Fig.~\ref{fig:2}d as a function of the delay. The peak in the graph corresponds to the highest anti-squeezing measured, which corresponds to the optimal overlap between squeezed light and LO.\\

\subsection{Acknowledgements}
The authors thank David Payne, Euan Allen, John Rarity and Patrick Yard for valuable scientific discussions and insights. This work was supported by the European Research Council starting grant ``PEQEM'' (ERC-2018-STG 803665), the EPSRC Fellowship (EP/M024385/1), the EPSRC grant ``Mono-Squeeze'' (EP/X016218/1), the Quantum Position, Navigation and Timing (QEPNT) Hub (EP/Z533178/1), the Quantum Sensing, Imaging and Timing (QuSIT) Hub (EP/Z533166/1), and the Integrated Quantum Networks (IQN) Hub (EP/Z533208/1). O.M.G. \& B.P. acknowledge support from EPSRC Quantum Engineering Centre for Doctoral Training (EP/S023607/1). A.S.C. acknowledges support from The Royal Society (URF/R/221019, RF/ERE/210098, RF/ERE/221060). J.C.F.M is grateful for support from his Philip Leverhulme Prize.\\

\subsection{Author contribution}
G.F conceived the experiment and designed the PIC and PCB. The pulse preparation scheme was designed by O.M.G and G.F. O.M.G constructed and characterised the setup , performed experiments and analysed results with input from B.P and R.N.C. R.A.H provided technic support for the laser system. E.C.R.D and O.M.G performed the JSI simulations and Schmidt mode analysis. A.S.C, J.C.F.M and G.F provided supervision for the project. O.M.G, R.N.C, A.S.C and G.F prepared the manuscript with input from all authors.


\bibliography{reference}

\end{document}